\documentclass[aps,pra,showpacs,twocolumn,floatfix,10pt]{revtex4-1}
\usepackage{graphicx}
\usepackage{amssymb}
\usepackage{subfigure}
\usepackage{amsmath}

\begin{document}

\title{Angular-momentum couplings in long-range Rb$_2$ Rydberg molecules}
\author{D.~A.~Anderson}
\author{S.~A.~Miller}
\author{G.~Raithel}
\affiliation{Department of Physics, University of Michigan, Ann Arbor, MI 48109}

\date{\today }
\begin{abstract}
We study angular-momentum couplings in $^{87}$Rb$_2$ Rydberg molecules formed between Rydberg and 5S$_{1/2}$ ground-state atoms.  We use a Fermi model that includes S-wave and P-wave singlet and triplet scattering of the Rydberg electron with the 5S$_{1/2}$ atom, along with the fine structure coupling of the Rydberg atom and hyperfine structure coupling of the 5S$_{1/2}$ atom.  We discuss the effects of these couplings on the adiabatic molecular potentials. We obtain bound-state energies, lifetimes, and electric and magnetic dipole moments for the vibrational ground states of the $^{87}$Rb$(n$D$+5$S$_{1/2})$ molecules in all adiabatic potentials, with fine and hyperfine structure included.  We also study the effect of the hyperfine structure on the deep $^3$S-wave- and $^3$P-wave-dominated adiabatic molecular potentials, which support high-$\ell$ $^{87}$Rb$_2$ Rydberg molecules.
\end{abstract}


\maketitle

\section{Introduction}
Rydberg molecules formed by the low-energy scattering of a Rydberg electron and a ground-state atom constitute a distinct class of molecular states that have in recent years become the subject of significant theoretical and experimental interest.  Developments include studies of diatomic Rydberg molecules in low-angular-momentum Rydberg S-states~\cite{Bendkowsky.2009,Tallant.2012}, P-states~\cite{Bellos.2013}, and D-states~\cite{Anderson.2014,Krupp.2014}, the realization of coherent bonding and dissociation of S-type molecules~\cite{Butscher.2010}, and the first observation of a permanent electric dipole moment in a homonuclear molecule~\cite{Li.2011}.  Polyatomic Rydberg molecules have also been generated~\cite{Bendkowsky.2010} and employed in a demonstration of the continuous transition between a few-body to many-body regime in an ultracold quantum gas~\cite{Gaj.2014}.

The theoretical framework for these molecules is generally well-established.  The interaction between a low-energy Rydberg electron and ground-state atom can be described using a Fermi pseudo-potential approach~\cite{Fermi.1934,Omont.1977,Greene.2000}.  In the Fermi model, the ground-state atom is treated as a delta-function perturber of the Rydberg-electron wave function, resulting in oscillatory potential curves with localized minima capable of sustaining bound molecular states.  In alkali systems, the potential curves are strongly affected by low-energy $^3$S-wave and $^3$P-wave electron-atom scattering resonances~\cite{Fabrikant.1986,Bahrim.2000,Bahrim.2001}.  The influence of these scattering resonances on the potentials has been studied in Rb$_2$. The $^3$S interaction leads to so-called ``trilobite'' molecules~\cite{Greene.2000}, which are very long-range (their size is on the order of $n^2$, where $n$ is the principal quantum number of the Rydberg level). The $^3$P interaction produces  potentials~\cite{Hamilton.2002} that are about an order of magnitude deeper and about a factor of five shorter-range than the ``trilobite'' potentials. The molecular potentials, wave functions and electric dipole moments for both Rb$_2$ and Cs$_2$ have also been calculated using a Green's function approach, accounting for the finite size of the perturbing ground-state atom via an effective short-range electron-atom interaction potential~\cite{Khuskivadze.2002,Chibisov.2002}.

In addition to the electron-atom scattering interaction, the molecular potentials and properties of long-range Rydberg molecules are dependent on the Rydberg-atom wave function as well as the angular-momentum couplings of the Rydberg- and ground-state constituents.  Rydberg molecules exhibit a broad range of different angular momentum coupling schemes.  For low-$\ell$ Rydberg molecules ($\ell\lesssim 2$ in rubidium), the angular-momentum coupling configurations span three Hund's cases [(a), (b), and (c)], dictated by the relative strength of the Rydberg atom's fine structure coupling compared to that of the $e^- + 5S_{1/2}$ scattering interaction.  The Rb$(nD_{j}+5S_{1/2})$ molecules are unique among the low-$\ell$ molecules because their scattering and fine-structure couplings are comparable, and fall anywhere between two Hund's cases (a) and (c) by a mere change in principal quantum number $n$.  For high-$\ell$ Rydberg molecules, the molecular binding interaction is stronger than the fine structure coupling, and is comparable to the hyperfine coupling of the ground-state perturber.  Inclusion of the 5S$_{1/2}$ hyperfine coupling in the model generates additional adiabatic potentials of mixed triplet and singlet character.  For both high- and low-$\ell$ molecules, the hyperfine structure results in additional adiabatic potentials deep enough to sustain bound states.

In this work we present a study on the influence of angular-momentum couplings on the properties of long-range Rydberg molecules.  We first describe a Fermi model for Rydberg molecules with the relevant angular-momentum couplings included. In our analysis we include singlet and triplet S- and P-wave scattering.  By selectively enabling the different interactions in the model, the effects of each of these individual interactions on the adiabatic molecular potentials are revealed.  Typical vibrational-state wave functions, binding energies, lifetimes, and dipole moments are discussed with an emphasis on the role of angular-momentum couplings.  We conclude with describing hyperfine effects on the relatively deeply bound ``trilobite'' Rydberg molecules~\cite{Greene.2000}.

\section{Fermi model}
\begin{figure}[htp]
\includegraphics[width=8.5cm]{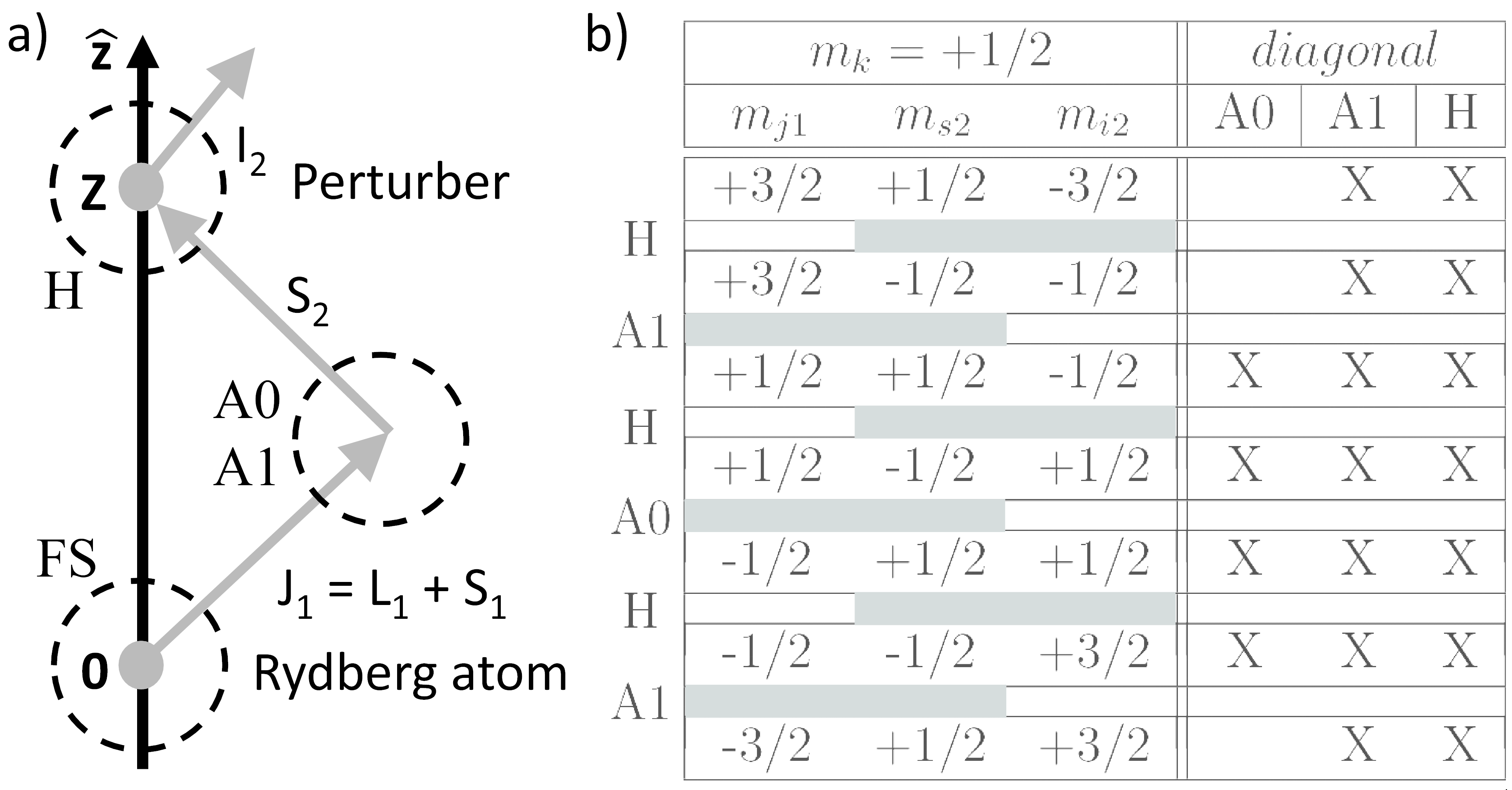}
\caption{a) Angular momentum coupling scheme for diatomic $^{87}$Rb$(nD_j+5S_{1/2}, F)$ Rydberg molecules.  The relevant interactions are circled.  Here, A0 and A1 denote $e^- + 5S_{1/2}$ scattering interactions involving the $m_{l1}=0$ (S-wave and P-wave) and the $\vert m_{l1} \vert =1$ (P-wave only) components of the Rydberg electron's state, respectively, and H denotes the hyperfine interaction of the $5S_{1/2}$ atom.  FS denotes the fine structure coupling.  b) States in the $m_k=m_{j1}+m_{s2}+m_{i2}=+1/2$ subspace and their relevant interactions.  In the left column, horizontal gray bars are placed between $(m_{j1}~m_{s2})$ or $(m_{s2}~m_{i2})$ for states in neighboring rows that are coupled by either the scattering or the hyperfine interaction.  The Xs in the right column indicate the interactions that have diagonal terms in $(m_{j1}~m_{s2}~m_{i2})$.}
\label{fig:coupling}
\end{figure}

We describe diatomic Rydberg molecules with a Fermi model~\cite{Fermi.1934,Omont.1977} taking into account the angular-momentum couplings in the Rydberg atom and perturbing ground-state atom system whose strengths are comparable to the Rydberg~$e^-$+perturber interaction.  A schematic of the relevant couplings is shown in Fig.~\ref{fig:coupling}a.  The perturbing $^{87}$Rb $5S_{1/2}$ atom is located at a position $Z$ from the ionic core of the Rydberg atom, which is fixed at the origin.  The internuclear axis is along $\hat{\bf{z}}$.  The orbital and spin angular momenta of the Rydberg atom are denoted by $L_1$ and $S_1$, respectively. For $L_1\leq2$ and within the $n$-range of interest, the Rydberg-atom fine structure is of the same order as the $e^- + 5S_{1/2}$ scattering interaction and is therefore included. The hyperfine coupling of the electron spin $S_2$ and nuclear spin $I_2$ of the perturbing $5S_{1/2}$ ground-state atom is also included, because it is several GHz and is, in most cases, stronger than the Rydberg electron's fine structure coupling and the $e^- + 5S_{1/2}$ scattering interaction. The orbital angular momentum of the $5S_{1/2}$ atom is $L_2=0$. The Rydberg atom's hyperfine structure decreases as $n^{-3}$; for the lowest $S$-states relevant to our work it does not exceed several~MHz~\cite{Meschede1987,Tauschinsky2013}, and it is much lower for higher-$\ell$ states~\cite{Sassmannshausen2013}.  The Rydberg atom's hyperfine structure is therefore not included.

For a Rydberg atom with its ionic core at the origin and the Rydberg electron located at $\bf r$, and with a $5S_{1/2}$ atom located at ${\bf{R}} = Z {\hat{\bf{z}}}$ , the Hamiltonian is written as

\begin{align}\label{Hamiltonian}
\hat{H}({\bf r},Z)&=\hat{H}_0 + \sum\limits_{i=S,T} 2 \pi A_{s}^{i}(k) \delta^3 ({\bf r} - Z {\hat{\bf{z}}} )\hat{\mathbb{I}}_{i} \\
    &\qquad {} + \sum\limits_{i=S,T} 6 \pi A_{p}^{i}(k)\delta^3 ({\bf r} - Z {\hat{\bf{z}}} )\overleftarrow\nabla\cdot\overrightarrow\nabla~\hat{\mathbb{I}}_{i} \nonumber \\
    &\qquad {} + A \hat{\bf S}_2 \cdot \hat{\bf I}_2 \nonumber
\end{align}

In the unperturbed Rydberg Hamiltonian $\hat{H}_0$ we use published quantum defects~\cite{Gallagher}, which account for core penetration and fine structure. For $\ell \ge 5$ we use the fine structure correction of hydrogen as well as a small quantum defect to account for core polarization, $\delta_\ell=0.75 \alpha_D /\ell^5$~\cite{Gallagher}, with a dipolar polarizability for Rb$^+$ of $\alpha_D=9.023$~atomic units~\cite{Litzen.1970}.  The energy-dependent S-wave ($l=0$) and P-wave ($l=1$) scattering lengths ($A_s$ and $A_p$, respectively) have the general form $A_{l}(k)= - \tan\delta_{l}/k^{2l+1}$, where $\delta_{l}$ is the $l$- and energy-dependent scattering phase shift.  For the calculations in the present work we use non-relativistic scattering phase shifts $\delta_{l=0}$ and $\delta_{l=1}$ generously provided by I. I. Fabrikant based on~\cite{Khuskivadze.2002}.  The electron momentum is given by $k=\sqrt{-1/n^{*2}_{0} + 2/r}$ (atomic units) in the classically allowed range of the Rydberg electron and $k=0$ elsewhere. Here, $n^{*}_{0}$ is the effective Rydberg quantum number
of the level of interest. To account for configuration interactions, we employ basis sets $\{ \vert n, L_1, J_1, m_{j1} \rangle \otimes \vert m_{s2}, m_{i2} \rangle \}$ that include all Rydberg levels with effective quantum numbers $\vert n^* - n^{*}_{0} \vert \lesssim 2.5$.  For the Rydberg atom, all $L_1$, $J_1$ and $m_{j1}$ are included (as in~\cite{Sadeghpour.2013}), and for the perturber atom
all $m_{s2}$ and $m_{i2}$ are included. The singlet (S) and triplet (T) channels of the $e^- + 5S_{1/2}$ scattering interaction include projectors $\hat{\mathbb{I}}_{(S,T)}$, defined as $\hat{\mathbb{I}}_T=\hat{\bf S}_1 \cdot \hat{\bf S}_2 + \frac{3}{4}$, which has an eigenvalue of one (zero) for the triplet (singlet) states, and $\hat{\mathbb{I}}_S=\hat{\mathbb{I}}-\hat{\mathbb{I}}_T$.  The operators $\hat{\bf S}_1$ and $\hat{\bf S}_2$ are the spins of the Rydberg electron and $5S_{1/2}$ atom, respectively, and $\hat{\mathbb{I}}$ is the identity operator. The last term in Eq.~\ref{Hamiltonian} accounts for the hyperfine interaction of the perturber.  The $^{87}$Rb($5S_{1/2}$) perturber atom has nuclear spin $\hat{\bf I}_2$ with $I_2=3/2$, and hyperfine levels $F_{<}=1$ and $F_{>}=2$, with a hyperfine coupling parameter $A = h \times 6.8~$GHz$/F_{>} = h \times 3.4$~GHz (in SI units).

Only levels with $m_{\ell 1}=0, \pm 1$ components have non-vanishing wave functions or wave function gradients on the internuclear axis. The S-wave interactions couple Rydberg states with $m_{\ell 1}=0$ components. The P-wave interactions couple states with $m_{\ell 1} =0$ components through the radial derivative of the Rydberg wave function, and states with $m_{\ell 1}=\pm 1$ components through the polar-angle derivative of the wave function. The electron scattering term in Eq.~\ref{Hamiltonian} conserves $m_{j1}+m_{s2}$, while the hyperfine term conserves $m_{s2}+m_{i2}$. The full Hamiltonian conserves $m_k : = m_{j1}+m_{s2}+m_{i2}$. Hence, the Hilbert space can be broken up into subspaces of fixed quantum number $m_k$. As an example, the subspace for $m_k=+1/2$ and its couplings via the S-wave, P-wave, and hyperfine interactions is shown in Fig.~\ref{fig:coupling}b. The gray bars in Fig.~\ref{fig:coupling}b illustrate that the couplings are organized in a block-diagonal structure in the magnetic quantum numbers.  Since the couplings via the scattering terms require $\vert m_{j1} \vert \leq 3/2$, the $^{87}$Rb$_2$ Rydberg molecules have $\vert m_k \vert \leq 7/2$.

\section{Adiabatic potentials}
\begin{figure}[htp]
\includegraphics[width=8.5cm]{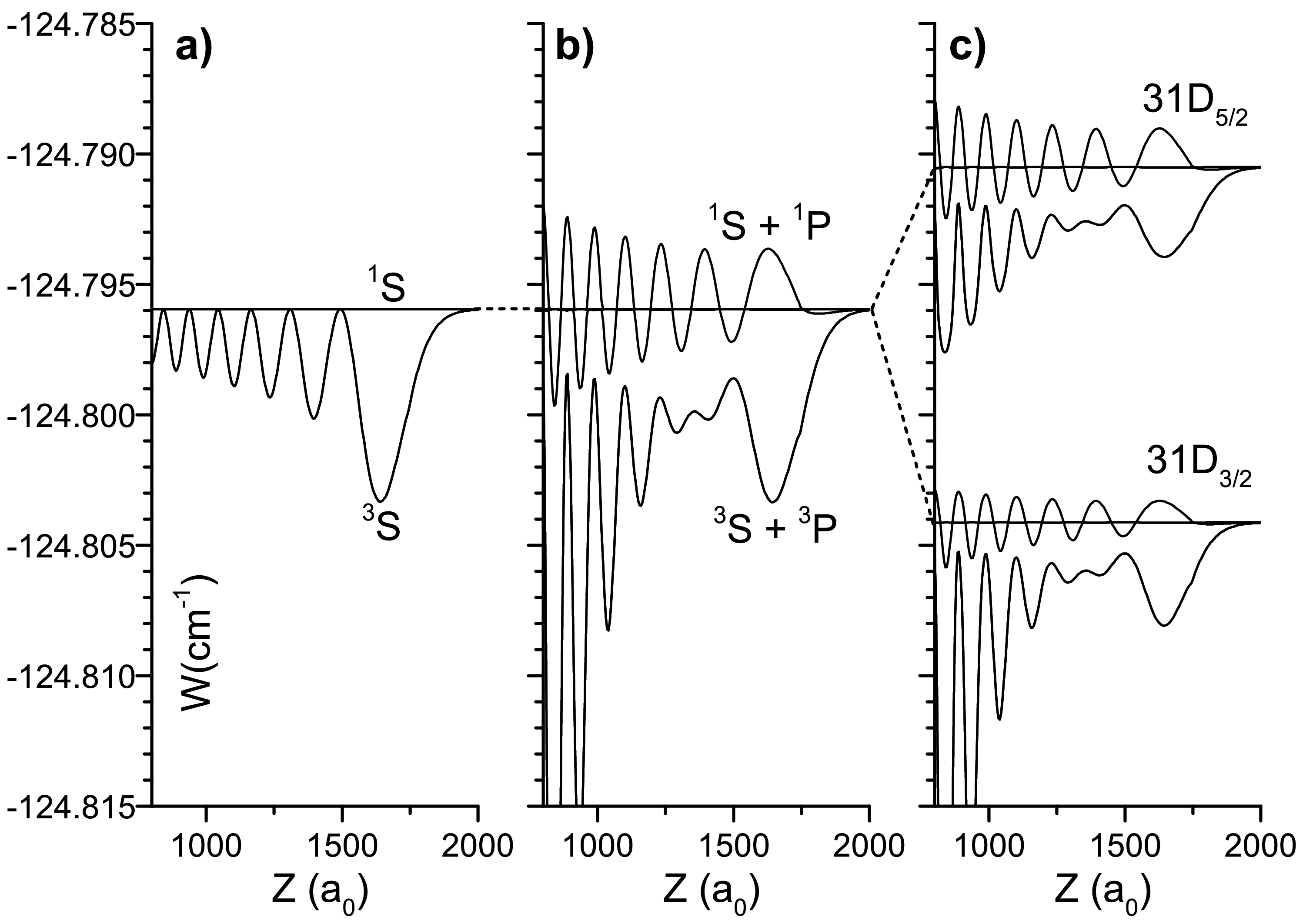}
\caption{Adiabatic potentials for the $31D+5S_{1/2}$ molecule with the following interaction terms in Eq.~\ref{Hamiltonian} selectively turned on (without the hyperfine interaction): a) $^3$S scattering, b) $^3$S, $^1$S, $^3$P, and $^1$P scattering, and c) $^3$S, $^1$S, $^3$P, and $^1$P scattering with fine structure coupling.}
\label{fig:terms}
\end{figure}

\begin{figure}[htp]
\includegraphics[width=8.5cm]{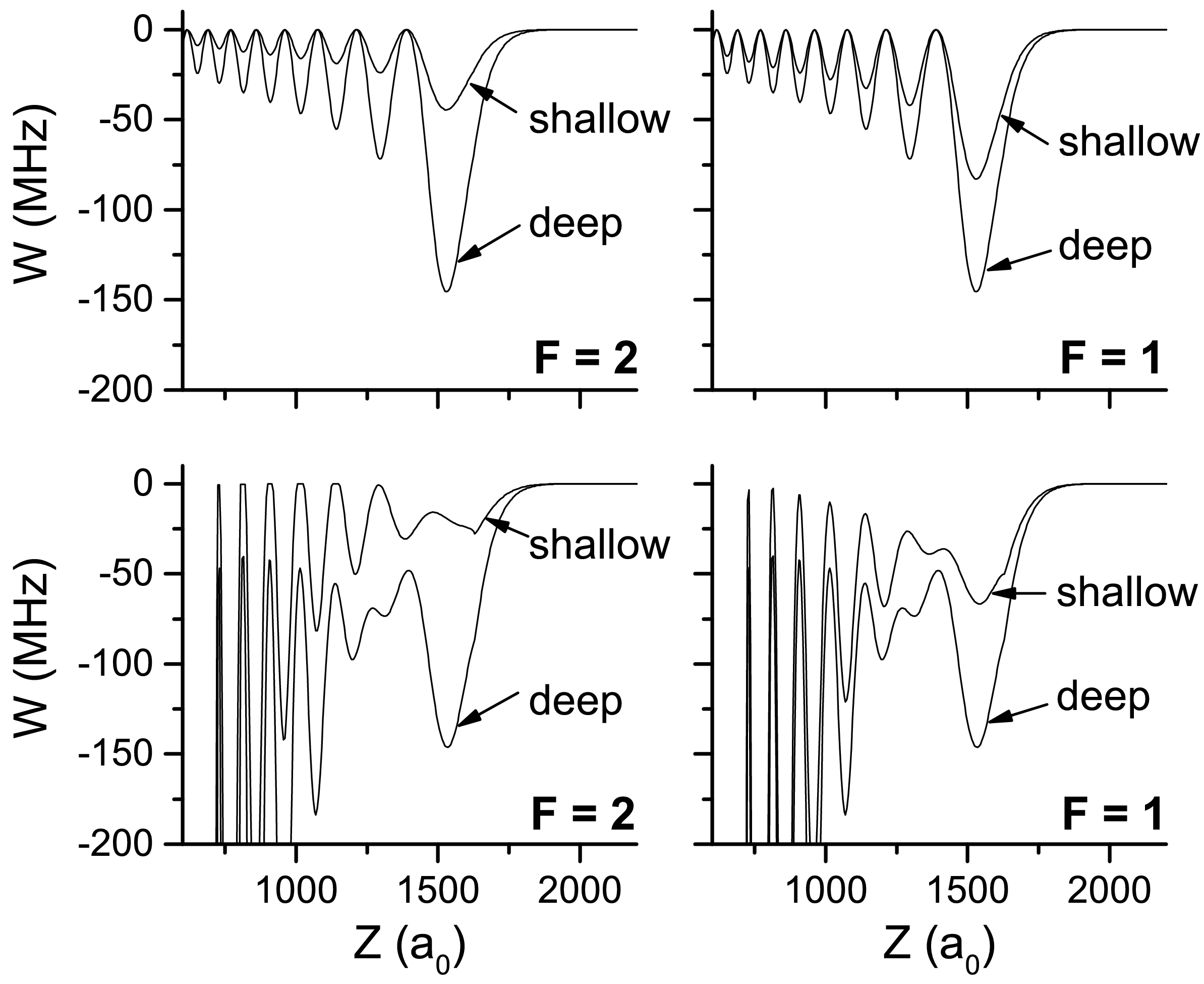}
\caption{Binding adiabatic potentials for the $^{87}$Rb$(30D_{3/2}+5S_{1/2}, F=1,2)$ molecules with fine and hyperfine structure included, with $^3$S scattering only (top row) and with $^3$S, $^1$S, $^3$P, and $^1$P scattering (bottom row). The hyperfine coupling leads to the shallow adiabatic potentials. The shallow potentials are different for the $F=1$ and $F=2$ hyperfine levels. The deep potentials do not depend on the hyperfine structure.}
\label{fig:terms2}
\end{figure}

Adiabatic molecular potentials $V_{i}(Z)$ (i is an arbitrary label) are obtained by diagonalizing the Hamiltonian in Eq.~\ref{Hamiltonian} for a grid of perturber atom positions, $Z$, through the extent of the Rydberg wave function.  To highlight the effects of the different terms in Eq.~\ref{Hamiltonian} on $V_{i}(Z)$, in Figs.~\ref{fig:terms} and ~\ref{fig:terms2} we show adiabatic potentials calculated for D-type Rydberg molecules with the different interaction terms in Eq.~\ref{Hamiltonian} selectively turned on.

First, we consider adiabatic potentials without hyperfine coupling.  Figure~\ref{fig:terms}a shows adiabatic potentials for the $31D+5S_{1/2}$ molecule with only the $^3$S interaction turned on and no fine structure coupling.  This results in three degenerate oscillatory potentials, one corresponding to each triplet state, and a flat potential corresponding to the singlet state (which has no $^3$S interaction).  The triplet potential curves are similar to those calculated in~\cite{Greene.2000}, in which the S-wave scattering length $ -\tan\delta_{s}/k$ is taken to first order in the electron momentum~\cite{Omont.1977}.  With an appropriate choice of the zero-energy S-wave scattering length, the $^3$S interaction reproduces measured binding energies of vibrational ground states of S-type Rydberg molecules~\cite{Bendkowsky.2009}.  Similarly, the $^3$S interaction with the addition of the fine structure reproduces vibrational ground states of D-type Rydberg molecules~\cite{Anderson.2014}. The effect of the fine structure coupling on the molecules is discussed further below.

Figure~\ref{fig:terms}b shows adiabatic potentials resulting from the $^3$S, $^1$S, $^3$P, and $^1$P scattering interactions turned on and no fine structure coupling.   At smaller internuclear separations, the influence of P-wave scattering becomes more significant due to higher electron energies closer to the Rydberg atom's ionic core (and therewith larger wave function gradients).  The effect of the $^3$P scattering interaction is evident in Fig.~\ref{fig:terms}b, where the inner wells become notably deeper, while the outermost well remains relatively unaffected.  The increasing contribution of P-wave scattering at smaller $Z$ generates deep molecular potentials.  The repulsive $^1$S and attractive $^1$P scattering interactions turn the flat singlet potential in Fig.~\ref{fig:terms}a into an oscillatory singlet potential with maxima above and wells below the dissociation threshold, as seen in Fig.~\ref{fig:terms}b.

Figure~\ref{fig:terms}c shows adiabatic potentials resulting from the the addition of the fine structure coupling of the Rydberg atom to the $^3$S, $^1$S, $^3$P, and $^1$P scattering interactions.  The top and bottom plots show $31D_j+5S_{1/2}$ potentials for the $j=5/2$ and 3/2 fine structure states, respectively. Qualitatively, the fine-structure coupling splits the molecular bonding strength (adiabatic-potential depth) of the fine-structure-free case between the two fine-structure levels, resulting in less deep potentials.  The splitting ratio depends on which Hund's case is more relevant (see Sec.~\ref{sec:Hund}).

The hyperfine interaction of the ground-state perturber in Eq.~\ref{Hamiltonian} mixes the singlet and some triplet scattering channels, resulting in the replacement of the pure singlet potentials in Fig.~\ref{fig:terms} with shallow adiabatic potentials of mixed singlet-triplet character. As an example, in Fig~\ref{fig:terms2} we show the adiabatic potentials for the $30D_{3/2}+5S_{1/2}, F=1,2$ molecules including the hyperfine interaction. The deep adiabatic potentials are of pure triplet character and are unaffected by the hyperfine interaction of the 5S$_{1/2}$ atom. The $F=1$ shallow, mixed singlet-triplet potentials are always deeper than the $F=2$ shallow potentials. The shallow potentials can typically sustain a few bound states that should be observable in experiments.

\section{Quasi-bound molecular states and lifetimes}
\begin{figure}[htp]
\includegraphics[width=8.5cm]{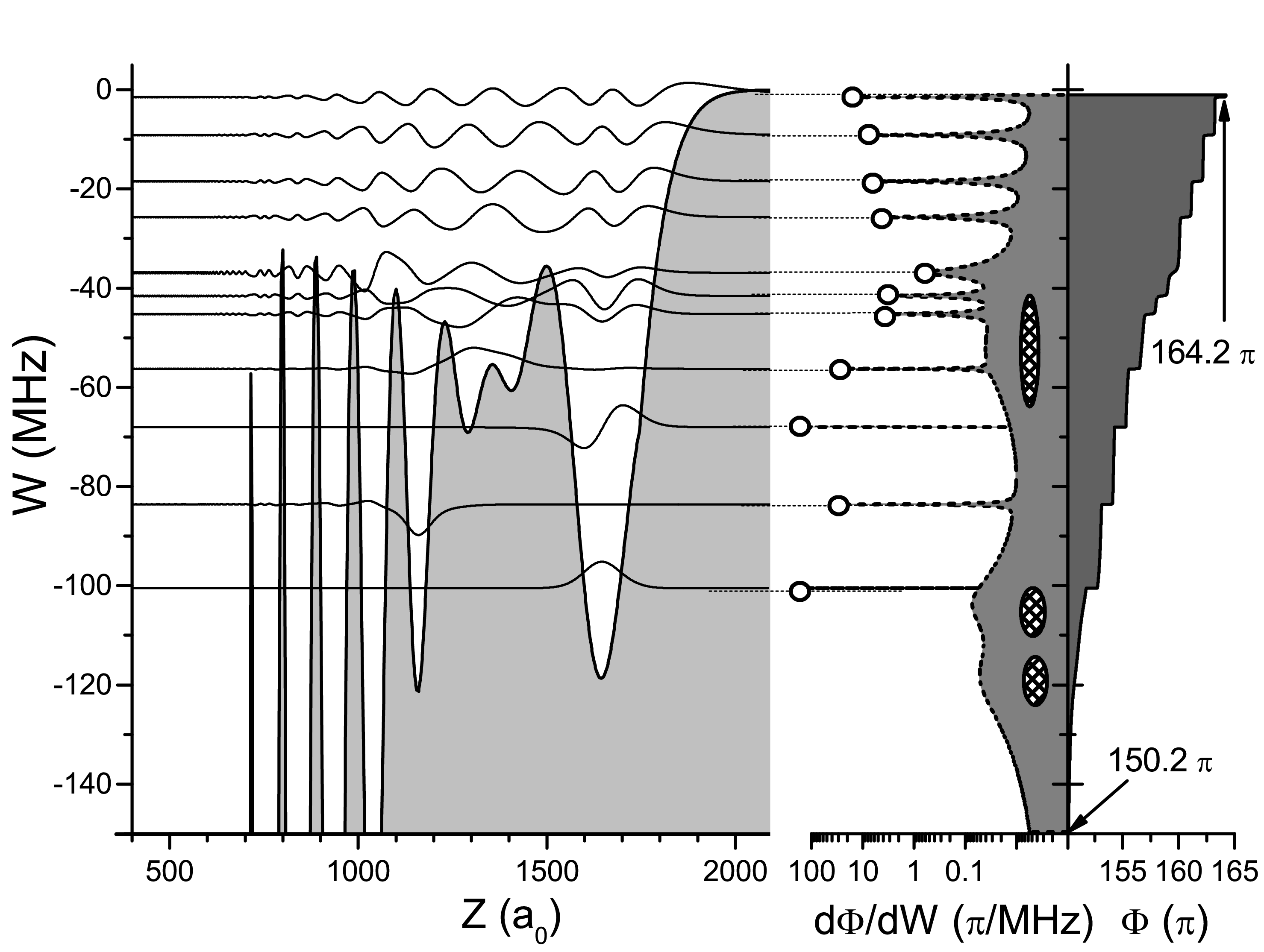}
\caption{(Left) Adiabatic potential of the $^{87}$Rb$(31D_{3/2}+5S_{1/2},F=2)$ Rydberg molecule and vibrational wave functions.  Each wave function corresponds to a narrow scattering resonance, characterized by a sudden change in the wave function phase by $\pi$ in the unbound, inner region of the potential. (Right) Wave function phase at location $Z=300~a_0$. (Middle) The maxima of $d\Phi / dW$, indicated by circles, are used to determine resonance widths and lifetimes. Several broad resonances (hatched regions) are spread out over the displayed energy range.}
\label{fig:31D1p5}
\end{figure}

Due to the lack of an inner potential barrier, the molecular vibrational states have the character of metastable scattering resonances in potentials that are unbound on the inside.  Figure~\ref{fig:31D1p5} shows the (deep) adiabatic potential for the $^{87}$Rb$(31D_{3/2}+5S_{1/2},F=2)$ molecule and its quasi-bound states.  These are qualitatively similar to those of S-type molecules~\cite{Bendkowsky.2010}.  The molecular wave functions consist of low-amplitude standing waves formed by outgoing and reflected ingoing waves in the region $Z \lesssim 1000~a_0$, and a high-amplitude, resonantly enhanced portion in the outer adiabatic-potential wells at $Z \sim 1500~a_0$. The latter are identified with quasi-bound molecular vibrational states that are metastable against tunneling-induced decay (decay into the region $Z \lesssim 1000~a_0$).

The resonances are found by computing the wave function phase at a fixed location in the unbound region (we use $Z=300~a_0$) as a function of energy $W$. The phase and its derivative are plotted as a function of energy in the right and middle panels of Fig.~\ref{fig:31D1p5}, respectively. The quasi-bound molecular states occur at energies at which the phase undergoes sudden changes of $\Delta \Phi=\pi$. The quasi-bound molecular states are centered at energies at which the derivative of the phase is maximal (circles in the middle panel of Fig.~\ref{fig:31D1p5}). The resonances obey a single-level Breit-Wigner formula with frequency linewidths of the quasi-bound molecular states given by $\Gamma_{\nu}= 2 /(h\times d\Phi/dW)$~\cite{Sakurai.Modern}, corresponding to lifetimes of $\tau = (\hbar/2) d\Phi/dW$.  The lifetimes scale as the Wigner tunneling time delay~\cite{Wigner.1955}. Resonances with larger slopes $d\Phi/dW$ in Fig.~\ref{fig:31D1p5} correspond to longer-lived quasi-bound states.  In addition to the sharp resonances there are broad resonances, as indicated by hatched regions in the middle panel of Fig.~\ref{fig:31D1p5}. While the broad resonances are not likely to cause observable effects in molecular spectra, they add to the total phase change over an energy range. For instance, within the energy range displayed in Fig.~\ref{fig:31D1p5} the broad resonances account for a phase change of 3$\pi$ and the quasi-bound molecular states for a change of 11$\pi$, corresponding to a total change of 14$\pi$. The distinction between broad and narrow resonances may, in practice, depend on experimental parameters such as excitation bandwidth and atom temperature (which affects Frank-Condon factors).

​\begin{table}[htp]
\caption{For the $^{87}$Rb$(31D_{3/2}+5S_{1/2},F=2)$ Rydberg molecule, we show the vibrational quantum number $\nu$, binding energy, linewidth $\Gamma_{\nu}$,  decay time, and bond length.}\label{lifetimetable}
\begin{ruledtabular}
\begin{tabular}{c|c|c|c|c}
$\nu$       & Energy  & Linewidth &Decay time& $<$Z$>$ \\
~           & (MHz)   & (MHz)     & ($\mu$s) &  (a$_0$)\\ \hline
0           & -100.52 & 1.79E-11  & 8.89E9   & 1647.5  \\
1           & -83.59  & 2.18E-2   & 7.30     & 1160.0  \\
2           & -68.03  & 6.00E-8   & 2.65E6   & 1655.3  \\
3           & -56.27  & 2.37E-2   & 6.72     & 1324.0  \\
4           & -45.24  & 1.48E-1   & 1.08     & 1413.0  \\
5           & -41.58  & 1.87E-1   & 0.85     & 1527.9  \\
6           & -36.91  & 1.01E+0   & 0.16     & 1169.3  \\
7           & -26.58  & 1.28E-1   & 1.24     & 1428.4  \\
8           & -18.48  & 9.99E-2   & 1.59     & 1469.0  \\
9           & -9.14   & 7.78E-2   & 2.05     & 1481.8  \\
10          & -1.48   & 3.71E-2   & 4.29      & 1697.0 \\\hline
\end{tabular}
\end{ruledtabular}
\end{table}​

In Table~\ref{lifetimetable} we give the binding energies, linewidths, decay times, and average internuclear separation of the scattering resonances shown in Fig.~\ref{fig:31D1p5}.  Each quasi-bound state is assigned a vibrational quantum number $\nu$, sequentially increasing from 0 for the most strongly bound state in the outermost well to 10 for the most weakly bound state.  Generally, one expects bound states furthest from the dissociation threshold to have the longest lifetimes.  Here, the ground and first excited states of the outermost potential well (labeled $\nu$=0 and 2, respectively) are well-confined and have lifetimes in the range of hours and seconds, respectively.  These lifetimes only reflect tunneling-induced decay.  The actual lifetimes of these molecular states are, in fact, much shorter due to additional decay mechanisms, such as radiative decay of the Rydberg state and collisions with ground-state atoms~\cite{Butscher.2011}.  An early dissociation process via the energy exchange between the Rydberg electron and ground-state atom has also been used to explain shorter molecular lifetimes observed in experiments~\cite{Junginger.2013}.  Nevertheless, the relatively long lifetimes and large Frank-Condon factors associated with these outermost states make them the easiest to isolate experimentally~\cite{Bendkowsky.2009,Anderson.2014,Gaj.2014}.  There are also several resonances in the inner potential wells ($\nu$=1 and 3) with lifetimes on the order of that of the Rydberg atom (which for $^{87}$Rb $31D_{3/2}$ is about 20~$\mu$s in a 300~K black-body radiation field).  The resonances at higher energies ($\nu=4-6$) exhibit shorter lifetimes because of the smaller potential barrier through which they more readily tunnel inward.  Surprisingly, above these short-lived resonances additional resonances with longer lifetimes appear ($\nu=7-10$).  Here, the inner oscillatory wells act like an aperiodic Bragg reflector of the molecular wave functions, resulting in unexpectedly long-lived resonances near the dissociation threshold.  The lifetime of these states is largest when a Bragg reflection condition is met. This occurs in a range of Rydberg principal quantum numbers $n$ at which the periodicity of the vibrational wave function approximately equals that of the Rydberg-electron wave function.  This Bragg-reflection has previously been described as an internal quantum reflection process in S-type molecules~\cite{Bendkowsky.2010}.

\section{Hund's cases for \MakeLowercase{n}D Rydberg molecules}
\label{sec:Hund}

Hund's coupling cases are widely used for classification of angular momentum couplings in diatomic molecules~\citep{Brown}.  Low-$\ell$ diatomic Rydberg molecules ($\ell\lesssim 2$ in rubidium) exhibit a variety of coupling cases, determined by the relative strength of the Rydberg atom's fine structure coupling to the e$^-$+perturber scattering interaction.  For $^{87}$Rb$(nD_{j}+5S_{1/2})$ molecules the fine structure coupling is comparable to the scattering interaction strength.  Due to this, D-type molecules trend from Hund's case (c) at large $n$, where the fine structure coupling exceeds the scattering interaction strength, to Hund's case (a) for $n \lesssim 35$, where the scattering interaction strength exceeds the fine structure coupling.  In previous work, we observed $^{87}$Rb$(nD_j+5S_{1/2}, F=2)(\nu=0)$ molecules in transition between the two Hund's coupling cases (a) and (c)~\cite{Anderson.2014}.  In this section, we focus on the potentials and quasi-bound states of the $^{87}$Rb$(nD_{j}+5S_{1/2})$, $j=3/2$ and $5/2$  molecules in the two limiting Hund's cases and in the transition regime.

\begin{figure}[htp]
\includegraphics[width=8.5cm]{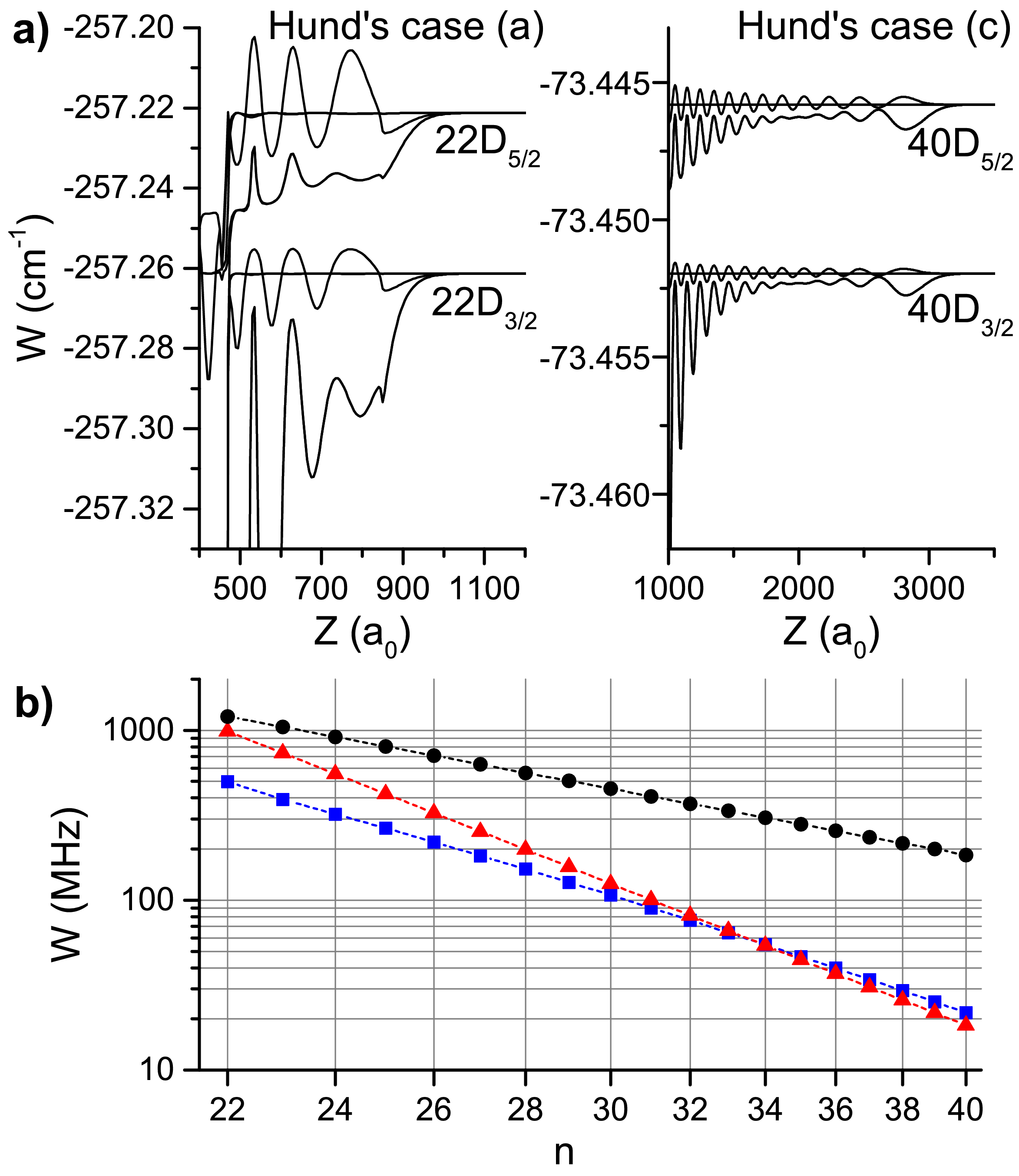}
\caption{a) Adiabatic potentials for $^{87}$Rb$(22D_{j}+5S_{1/2})$ (left) and $^{87}$Rb$(40D_{j}+5S_{1/2})$ (right) with $^3S, ^1S, ^3P,  ^1P$ interactions and no HFS interaction.  b) Binding energies for the $\nu = 0$ ground vibrational state of the $nD_{5/2}+5S_{1/2}$ (blue squares) and $nD_{3/2}+5S_{1/2}$ (red triangles) molecular potentials versus $n$.  The $D$ fine-structure splitting is also plotted (black circles).}
\label{fig:Hunds_bes}
\end{figure}

Figure~\ref{fig:Hunds_bes}a shows $V_{i}(Z)$ for the j=3/2 and 5/2 $(22D_j+5S_{1/2})$ molecules (left) and j=3/2 and 5/2 $(40D_j+5S_{1/2})$ molecules (right) calculated with all interaction terms in Eq.~\ref{Hamiltonian}, excluding hyperfine-structure coupling.  At high $n$, the molecules trend towards Hund's case (c), where the dominant adiabatic molecular potentials are reduced by the product of two Clebsch-Gordan coefficients of the type $\langle m_{\ell} = 0, m_{s}= \pm 1/2 \vert j, m_j= \pm 1/2 \rangle$.  This leads to adiabatic potentials whose depths carry spin-dependent factors $\ell / (2 \ell +1)$ for $j=\ell-1/2$ and $(\ell+1) / (2 \ell +1)$ for $j=\ell+1/2$~\cite{Sakurai.Modern}.  For $nD$ ($\ell=2$) molecules at high $n$, the depth ratio of the potentials for $j=3/2$ and $j=5/2$  approaches 2/3.  This is seen in the depths of the outermost wells of the high-$n$ j=3/2 and 5/2 $(40D_j+5S_{1/2})$ molecular potentials in Fig.~\ref{fig:Hunds_bes}a.  For decreasing $n$, the fine structure splitting increases as $n^{-3}$ while the scattering interaction strength increases as $n^{-6}$, and the molecules tend towards Hund's case (a).  The j=3/2 and 5/2 $(22D_j+5S_{1/2})$ molecular potentials in Fig.~\ref{fig:Hunds_bes} exhibit this case, where the scattering interaction strength is large compared to the fine structure splitting.  In the low-$n$ limit, the $j=3/2$ potential becomes notably deeper than the $j=5/2$ potential.

The binding energies of the vibrational ground states in the outermost potential wells, $W_{i,\nu=0}$, closely track the depth of those wells and therefore mirror the transition of the molecule between the two Hund's cases. In Fig.~\ref{fig:Hunds_bes}b, we plot $W_{i,\nu=0}$ for the j=3/2 and 5/2 $(nD_{j}+5S_{1/2})$ molecules for a range of $n$ from Hund's case (a) (low-$n$) to Hund's case (c) (high-$n$).  In the high-$n$ limit, the molecular binding energies $W_{i,\nu=0}$ for both fine structure levels approximately scale as $n^{-6}$, inversely with the atomic volume.  In the low-$n$ limit, the lowest binding energy for the lower fine-structure level (j=3/2) is larger and continues to scale as $\sim n^{-6}$, while that of the upper fine structure level (j=5/2) trends towards a $\sim n^{-3}$ scaling, approaching the scaling of the fine structure splitting.  For the Rb$(nD_{j}+5S_{1/2})$ molecules, the binding energies of the vibrational ground states in the outermost potential wells for j=3/2 and j=5/2 are approximately equal at $n=34$ (see Fig.~\ref{fig:Hunds_bes}b).  Further, due to the described Hund's case behavior, at sufficiently low $n$ the inner wells of adiabatic potentials of the $nD_{3/2}+5S_{1/2}$ molecule become deep enough to support vibrational states that are more deeply-bound and long-lived than the ground states of the outermost well of the $nD_{5/2}+5S_{1/2}$ potentials.

\section{Electric and magnetic dipole moments}

Rydberg molecules present the only known case of homonuclear molecules with permanent electric dipole moments~\cite{Greene.2000,Li.2011}.  For high-$\ell$ Rydberg molecules dipole moments on the order of $10^3~ea_0$ are predicted to exist~\cite{Greene.2000}.  Smaller permanent dipole moments arise in low-$\ell$ S-, P-, and D-type molecules from fractional admixing of high-$\ell$ state character.  Permanent electric dipole moments of $\sim 0.5~ea_0$ have previously been measured in rubidium S-type Rydberg molecules~\cite{Li.2011}.  Dipolar cesium Rydberg molecules with electric dipole moments of $\sim 5-50~ea_0$~\cite{Tallant.2012} have also been prepared.  In this section we calculate both the electric and magnetic dipole moments for $nD$ Rydberg molecules, with all terms in the Hamiltonian in Eq.~1 included.

\begin{figure*}[htp]
\includegraphics[width=22.5cm]{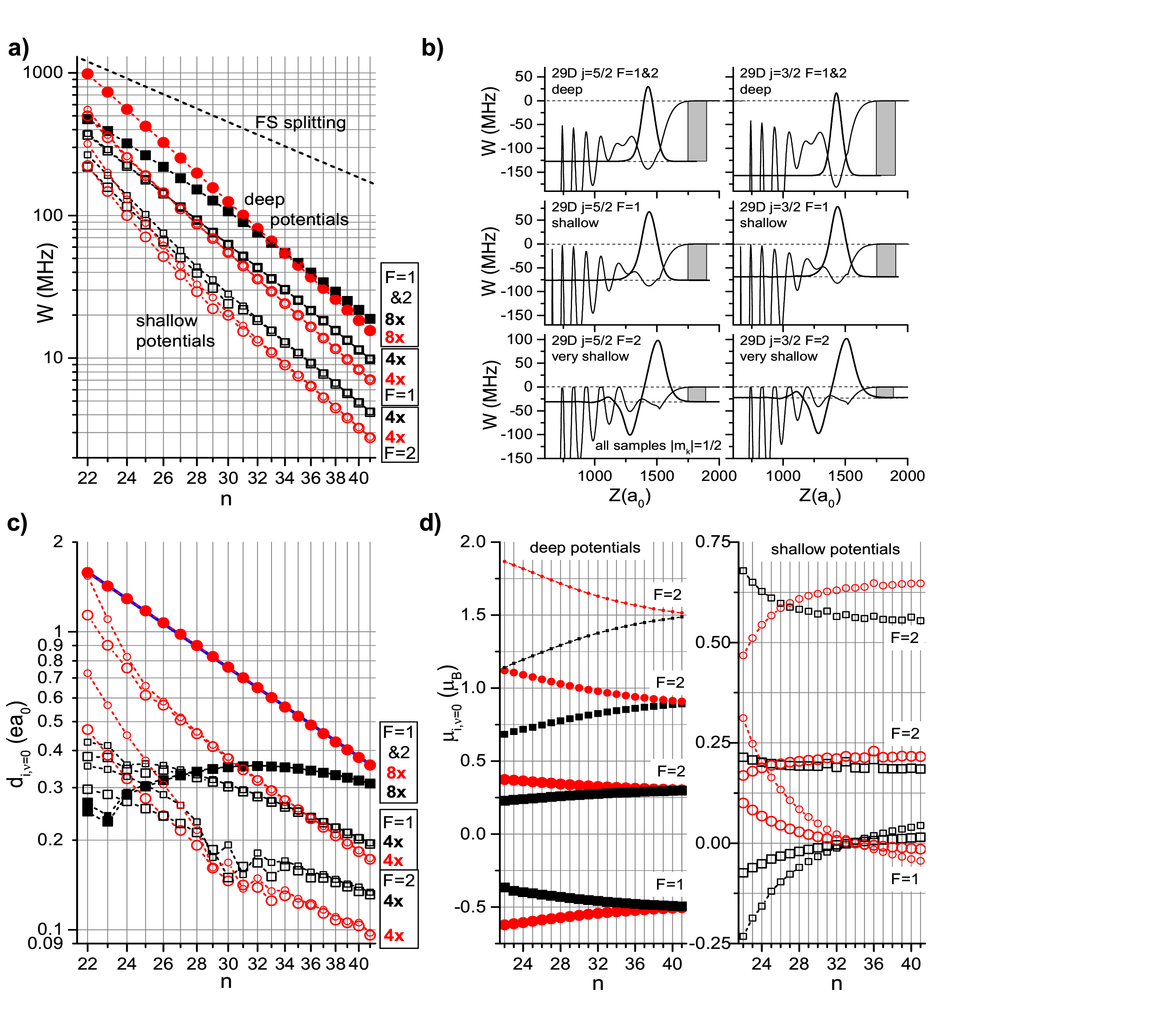}
\caption{a) Binding energies for $\nu=0$ in the outermost potential wells vs $n$, with all terms in Eq.~1 included. Symbol legend: Solid: deep, triplet potentials. These are identical for $F=1$ and $F=2$. Hollow: shallow, mixed singlet-triplet potentials. These are different for $F=1$ and $F=2$. Black squares: $j=5/2$. Red circles: $j=3/2$. Large size:
$\vert m_k \vert= 1/2$. Medium size:
$\vert m_k \vert= 3/2$. Small size:
$\vert m_k \vert= 5/2$. The $n$D$_j$ fine structure splitting is plotted for reference. Numbers indicate degeneracies summed over all $m_k$. b) Representative adiabatic potentials and wave functions for $^{87}$Rb$(29D_{j}+5S_{1/2}, F=1,2)(\nu=0)$, for $j=5/2$ (left column) and $j=3/2$ (right column). The pure triplet potentials are the same for $F=1$ and $F=2$ (top row), while the mixed singlet/triplet potentials are generally shallow and different for $F=1$ (middle row) and F=2 (bottom row). The gray bars on the right indicate binding energy, to visualize that the deep potentials are closer to Hund's case (a) ($j=3/2$ potential deeper than $j=5/2$ potential) than the shallow ones ($j=5/2$ potentials deeper than $j=3/2$ potentials). c) Electric dipole moments $d_{i,\nu}$ for $\nu=0$ in the outermost potential wells vs $n$, with all terms in Eq.~1 included. Symbol legend as in panel a). The blue line through the data for the deep $j=3/2$ potentials is an allometric fit with exponent -2.4. d) Magnetic dipole moments $\mu_{i,\nu}$ for $\nu=0$ in the outermost potential wells vs $n$, with all terms in Eq.~1 included. Symbol legend as in panel a). We only show data for positive $m_k$ (results for negative  $m_k$  are the same with flipped sign). There are no degeneracies in  $\mu_{i,\nu=0}$.}
\label{fig:dipolemoments}
\end{figure*}

We obtain the adiabatic electric [$d_{i,z}(Z)$] and magnetic [$\mu_{i,z}(Z)$] dipole moments in the diagonalization of Eq.~\ref{Hamiltonian}.  The dipole moments of a molecular state $\nu$ follow from the expectation values of $d_{i,z}(Z)$ and $\mu_{i,z}(Z)$ over the vibrational wave function densities,

\begin{eqnarray}
  d_{i,\nu} &=& \int\vert \Psi_{i,\nu}(Z) \vert^2 d_{i,z}(Z)dZ\nonumber \\
  \mu_{i,\nu} &=& \int\vert \Psi_{i,\nu}(Z) \vert^2 \mu_{i,z}(Z)dZ\quad.
\end{eqnarray}

Electric dipole moments for the ground vibrational states of the j=3/2 and 5/2 $^{87}$Rb$(nD_{j}+5S_{1/2}, F=1,2)$ deep (pure triplet) and shallow (mixed singlet/triplet) molecular potentials are shown in Fig.~\ref{fig:dipolemoments}. In Figs.~\ref{fig:dipolemoments}a and b we show the binding energies and representative potentials with wave functions, respectively, for all angular-momentum coupling cases that arise from Eq.~1.  We note that for $^{87}$Rb$(nD_{j}+5S_{1/2}, F=1,2)$ molecules the hyperfine quantum numbers $F$ are well-defined because the hyperfine coupling is much larger than the molecular binding. As seen in Figs.~\ref{fig:dipolemoments}a and b, the delineation between Hund's cases (a) and (c) is shifted to lower $n$ for the shallow potentials (which are due to hyperfine-induced mixing of singlet and triplet states at the $5S_{1/2}$ atom).  Essentially, the generally weaker scattering interaction associated with the mixed singlet/triplet cases pushes those molecules more towards Hund's case (c).

The electric dipole moments of the deep molecular potentials are 8-fold degenerate when summed over all $m_k$, and those of the shallow potentials are 4-fold degenerate (see Fig.~\ref{fig:dipolemoments}c). It is noted that with decreasing $n$ the degeneracies in energy and electric dipole moment become increasingly lifted.  This may be attributed to a stronger configuration mixing at low $n$ caused by the relative increase of the $e^-$+perturber scattering term in Eq.~1.
The only case in which the dipole moments of the $^{87}$Rb$(nD_{j}+5S_{1/2}, F=1,2)$ molecules exhibit a clear scaling behavior in $n$ is for the $j=3/2$ deep potentials (which are the same for $F=1$ and $F=2$).  A fit to the
top-most data set in Fig.~\ref{fig:dipolemoments}c gives a $\propto n^{-2.4}$ scaling, similar to a result found previously for $^{87}$Rb$_2$ $S$-type Rydberg molecules~\cite{Li.2011}.  Here, the scattering-induced mixing between the atomic Rydberg levels gives rise to a significant change in the electric dipole moments as a function of $n$. In the range $n \gtrsim 30$, the electric dipole moments for both
types of shallow potentials of $^{87}$Rb$(nD_{3/2}+5S_{1/2}, F=1,2)$ also follow a scaling similar to $\propto n^{-2.4}$. The electric  dipole moments for the upper fine structure component $j=5/2$ have less clear scaling trends (squares in Fig.~\ref{fig:dipolemoments}c). In particular, the electric  dipole moments for the deep $^{87}$Rb$(nD_{5/2}+5S_{1/2}, F=1,2)$ potentials do not exhibit a clear scaling behavior. This is likely a result of the transition from Hund's case (a) to Hund's case (c).

The magnetic moments for the molecular states are shown in Fig.~\ref{fig:dipolemoments}d. The magnetic moments are non-degenerate due to the different $g$-factors of the involved types of spins. The $n$-dependence again reflects the transition in angular-momentum coupling behavior between the two Hund's cases (a) and (c). We expect that experiments in weak electric and magnetic fields can reveal the electric and magnetic dipole moments of the vibrational states. The dipole moments computed in this work are for the weak-field limit, {\sl i.e.} results are expected to be accurate as long as molecular Zeeman and Stark shifts are smaller than other relevant energy scales (such as the energy splitting between adjacent vibrational states).

\section{Hyperfine-structure effects in deep $^3$S- and $^3$P-dominated potentials}

\begin{figure}[htp]
\includegraphics[width=8.5cm]{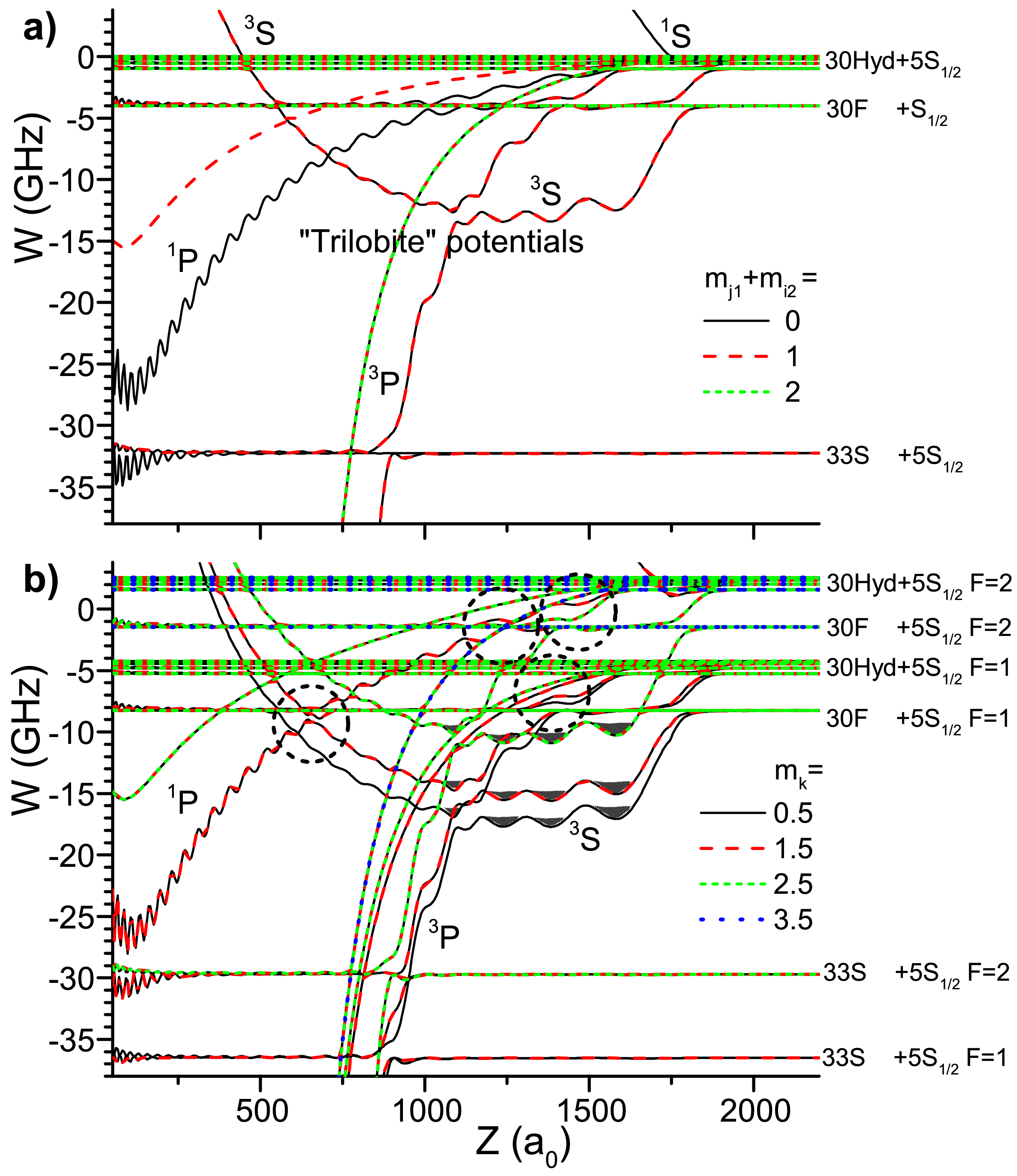}
\caption{High-$\ell$ adiabatic potentials near $n=30$ of $^{87}$Rb$_2$ Rydberg + 5S$_{1/2}$ molecules a) without and b) with hyperfine structure included. The plots indicate the ``trilobite'' potentials~\cite{Greene.2000}, the dominant types of scattering interactions leading to deep potentials, the asmyptotic states of the potentials, the regions where bound ``trilobite'' molecules may be found (gray areas), and the intersections between ``trilobite'' potentials and $F$ ($\ell=3$) lines (dashed circles).  }
\label{fig:trilobite}
\end{figure}

The ``trilobite'' molecules~\cite{Greene.2000} generated by the $^3$S interaction are long-range
and on the order of 10~GHz deep, which is on the same order as the hyperfine interaction of the $5S_{1/2}$ perturber.
It is therefore of interest to explore hyperfine effects on the  ``trilobite'' adiabatic potentials.  In Figs.~7a and b we show these potentials without and with hyperfine coupling for $n=30$, for all relevant values of $m_{j1} + m_{s2}$  and $m_k = m_{j1} + m_{s2} + m_{i2}$, respectively.  While the hyperfine interaction does not affect the general shape and depth of the ``trilobite'' potentials, with hyperfine structure included there are three instead of only one. Since the number of $^3$P potential curves also triples, the crossing pattern  between  $^3$S and  $^3$P potentials becomes considerably more complex, as seen in Fig.~\ref{fig:trilobite}. Considering that the modulations near the bottom of the ``trilobite'' potentials are several 100~MHz deep, which is sufficient to support individual bound vibrational states, the hyperfine structure is expected to have profound effects on the detailed vibrational level structure of these molecules.  In Fig.~7b, one may expect to find long-lived $^3$S-dominated states in the gray regions. Further, as indicated by the circles in Fig.~7b, the crossing locations and the detailed coupling behavior of $\ell$=3 Rydberg levels (which are optically accessible form low-lying atomic states) to molecular states in the long-range potentials also strongly depend on the hyperfine structure of the system under investigation.

\section{Conclusion}
In summary, we have systematically explored the role of fine-structure and hyperfine angular-momentum couplings in $^{87}$Rb$_2$ molecules formed between Rydberg and 5S$_{1/2}$ ground-state atoms.  As has been done extensively in previous work, we have treated the electron-5S$_{1/2}$ scattering with a Fermi model that includes S-wave and P-wave singlet and triplet scattering.  The fine structure mostly influences the behavior of low-$\ell$ Rydberg molecules.  We have explored in detail how $^{87}$Rb$(nD+5S_{1/2})$ molecules realize Hund's cases (a) and (c). The hyperfine structure originates in the perturber atom and therefore has consequences for all types (low-$\ell$ and high-$\ell$) of Rydberg molecules. In the case of $^{87}$Rb$(nD+5S_{1/2})$ molecules, mixing of singlet and triplet potentials results in a set of shallow adiabatic potentials, whose quasi-bound states should be experimentally observable (in addition to those in the hyperfine-independent pure triplet potentials).  We have obtained electric and magnetic dipole moments, which could, in future work, be measured spectroscopically in weak electric and magnetic fields.  The hyperfine structure has also been seen to alter the deep, $^3$S-scattering-induced trilobite potentials as well as their crossing behavior with $\ell$=3 Rydberg states; experimental spectroscopic studies should reveal these details.  Since molecular level energies and properties are very sensitive to the scattering phase shifts used in the Fermi model, we expect that spectroscopy of Rydberg molecules can serve as a sensitive tool to provide measurement-based input to future theoretical studies in low-energy electron scattering.

We thank I. I. Fabrikant for providing scattering phase shifts used in our calculations.  The work was supported by the AFOSR (FA9550-10-1-0453) and the NSF (PHY-1205559).

*anderda@umich.edu

\bibliographystyle{apsrev4-1}

\begin{thebibliography}{29}%
\makeatletter
\providecommand \@ifxundefined [1]{%
 \@ifx{#1\undefined}
}%
\providecommand \@ifnum [1]{%
 \ifnum #1\expandafter \@firstoftwo
 \else \expandafter \@secondoftwo
 \fi
}%
\providecommand \@ifx [1]{%
 \ifx #1\expandafter \@firstoftwo
 \else \expandafter \@secondoftwo
 \fi
}%
\providecommand \natexlab [1]{#1}%
\providecommand \enquote  [1]{``#1''}%
\providecommand \bibnamefont  [1]{#1}%
\providecommand \bibfnamefont [1]{#1}%
\providecommand \citenamefont [1]{#1}%
\providecommand \href@noop [0]{\@secondoftwo}%
\providecommand \href [0]{\begingroup \@sanitize@url \@href}%
\providecommand \@href[1]{\@@startlink{#1}\@@href}%
\providecommand \@@href[1]{\endgroup#1\@@endlink}%
\providecommand \@sanitize@url [0]{\catcode `\\12\catcode `\$12\catcode
  `\&12\catcode `\#12\catcode `\^12\catcode `\_12\catcode `\%12\relax}%
\providecommand \@@startlink[1]{}%
\providecommand \@@endlink[0]{}%
\providecommand \url  [0]{\begingroup\@sanitize@url \@url }%
\providecommand \@url [1]{\endgroup\@href {#1}{\urlprefix }}%
\providecommand \urlprefix  [0]{URL }%
\providecommand \Eprint [0]{\href }%
\providecommand \doibase [0]{http://dx.doi.org/}%
\providecommand \selectlanguage [0]{\@gobble}%
\providecommand \bibinfo  [0]{\@secondoftwo}%
\providecommand \bibfield  [0]{\@secondoftwo}%
\providecommand \translation [1]{[#1]}%
\providecommand \BibitemOpen [0]{}%
\providecommand \bibitemStop [0]{}%
\providecommand \bibitemNoStop [0]{.\EOS\space}%
\providecommand \EOS [0]{\spacefactor3000\relax}%
\providecommand \BibitemShut  [1]{\csname bibitem#1\endcsname}%
\let\auto@bib@innerbib\@empty
\bibitem [{\citenamefont {Bendkowsky}\ \emph {et~al.}(2009)\citenamefont
  {Bendkowsky}, \citenamefont {Butscher}, \citenamefont {Nipper}, \citenamefont
  {Shaffer}, \citenamefont {L\"ow},\ and\ \citenamefont
  {Pfau}}]{Bendkowsky.2009}%
  \BibitemOpen
  \bibfield  {author} {\bibinfo {author} {\bibfnamefont {V.}~\bibnamefont
  {Bendkowsky}}, \bibinfo {author} {\bibfnamefont {B.}~\bibnamefont
  {Butscher}}, \bibinfo {author} {\bibfnamefont {J.}~\bibnamefont {Nipper}},
  \bibinfo {author} {\bibfnamefont {J.~P.}\ \bibnamefont {Shaffer}}, \bibinfo
  {author} {\bibfnamefont {R.}~\bibnamefont {L\"ow}}, \ and\ \bibinfo {author}
  {\bibfnamefont {T.}~\bibnamefont {Pfau}},\ }\href {\doibase
  10.1038/nature07945} {\bibfield  {journal} {\bibinfo  {journal} {Nature}\
  }\textbf {\bibinfo {volume} {458}},\ \bibinfo {pages} {1005} (\bibinfo {year}
  {2009})}\BibitemShut {NoStop}%
\bibitem [{\citenamefont {Tallant}\ \emph {et~al.}(2012)\citenamefont
  {Tallant}, \citenamefont {Rittenhouse}, \citenamefont {Booth}, \citenamefont
  {Sadeghpour},\ and\ \citenamefont {Shaffer}}]{Tallant.2012}%
  \BibitemOpen
  \bibfield  {author} {\bibinfo {author} {\bibfnamefont {J.}~\bibnamefont
  {Tallant}}, \bibinfo {author} {\bibfnamefont {S.~T.}\ \bibnamefont
  {Rittenhouse}}, \bibinfo {author} {\bibfnamefont {D.}~\bibnamefont {Booth}},
  \bibinfo {author} {\bibfnamefont {H.~R.}\ \bibnamefont {Sadeghpour}}, \ and\
  \bibinfo {author} {\bibfnamefont {J.~P.}\ \bibnamefont {Shaffer}},\ }\href
  {\doibase 10.1103/PhysRevLett.109.173202} {\bibfield  {journal} {\bibinfo
  {journal} {Phys. Rev. Lett.}\ }\textbf {\bibinfo {volume} {109}},\ \bibinfo
  {pages} {173202} (\bibinfo {year} {2012})}\BibitemShut {NoStop}%
\bibitem [{\citenamefont {Bellos}\ \emph {et~al.}(2013)\citenamefont {Bellos},
  \citenamefont {Carollo}, \citenamefont {Banerjee}, \citenamefont {Eyler},
  \citenamefont {Gould},\ and\ \citenamefont {Stwalley}}]{Bellos.2013}%
  \BibitemOpen
  \bibfield  {author} {\bibinfo {author} {\bibfnamefont {M.~A.}\ \bibnamefont
  {Bellos}}, \bibinfo {author} {\bibfnamefont {R.}~\bibnamefont {Carollo}},
  \bibinfo {author} {\bibfnamefont {J.}~\bibnamefont {Banerjee}}, \bibinfo
  {author} {\bibfnamefont {E.~E.}\ \bibnamefont {Eyler}}, \bibinfo {author}
  {\bibfnamefont {P.~L.}\ \bibnamefont {Gould}}, \ and\ \bibinfo {author}
  {\bibfnamefont {W.~C.}\ \bibnamefont {Stwalley}},\ }\href {\doibase
  10.1103/PhysRevLett.111.053001} {\bibfield  {journal} {\bibinfo  {journal}
  {Phys. Rev. Lett.}\ }\textbf {\bibinfo {volume} {111}},\ \bibinfo {pages}
  {053001} (\bibinfo {year} {2013})}\BibitemShut {NoStop}%
\bibitem [{\citenamefont {Anderson}\ \emph {et~al.}(2014)\citenamefont
  {Anderson}, \citenamefont {Miller},\ and\ \citenamefont
  {Raithel}}]{Anderson.2014}%
  \BibitemOpen
  \bibfield  {author} {\bibinfo {author} {\bibfnamefont {D.~A.}~\bibnamefont
  {Anderson}}, \bibinfo {author} {\bibfnamefont {S.~A.}~\bibnamefont {Miller}}, \
  and\ \bibinfo {author} {\bibfnamefont {G.}~\bibnamefont {Raithel}},\ }\href
  {\doibase 10.1103/PhysRevLett.112.163201} {\bibfield  {journal} {\bibinfo
  {journal} {Phys. Rev. Lett.}\ }\textbf {\bibinfo {volume} {112}},\ \bibinfo
  {pages} {163201} (\bibinfo {year} {2014})}\BibitemShut {NoStop}%
\bibitem [{\citenamefont {Krupp}\ \emph {et~al.}(2014)\citenamefont {Krupp},
  \citenamefont {Gaj}, \citenamefont {Balewski}, \citenamefont {Ilzh\"ofer},
  \citenamefont {Hofferberth}, \citenamefont {L\"ow}, \citenamefont {Pfau},
  \citenamefont {Kurz},\ and\ \citenamefont {Schmelcher}}]{Krupp.2014}%
  \BibitemOpen
  \bibfield  {author} {\bibinfo {author} {\bibfnamefont {A.~T.}~\bibnamefont
  {Krupp}}, \bibinfo {author} {\bibfnamefont {A.}~\bibnamefont {Gaj}}, \bibinfo
  {author} {\bibfnamefont {J.~B.}~\bibnamefont {Balewski}}, \bibinfo {author}
  {\bibfnamefont {P.}~\bibnamefont {Ilzh\"ofer}}, \bibinfo {author}
  {\bibfnamefont {S.}~\bibnamefont {Hofferberth}}, \bibinfo {author}
  {\bibfnamefont {R.}~\bibnamefont {L\"ow}}, \bibinfo {author} {\bibfnamefont
  {T.}~\bibnamefont {Pfau}}, \bibinfo {author} {\bibfnamefont {M.}~\bibnamefont
  {Kurz}}, \ and\ \bibinfo {author} {\bibfnamefont {P.}~\bibnamefont
  {Schmelcher}},\ }\href {\doibase 10.1103/PhysRevLett.112.143008} {\bibfield
  {journal} {\bibinfo  {journal} {Phys. Rev. Lett.}\ }\textbf {\bibinfo
  {volume} {112}},\ \bibinfo {pages} {143008} (\bibinfo {year}
  {2014})}\BibitemShut {NoStop}%
\bibitem [{\citenamefont {Butscher}\ \emph {et~al.}(2010)\citenamefont
  {Butscher}, \citenamefont {Nipper}, \citenamefont {Balewski}, \citenamefont
  {Kukota}, \citenamefont {Bendkowsky}, \citenamefont {L\"ow},\ and\
  \citenamefont {Pfau}}]{Butscher.2010}%
  \BibitemOpen
  \bibfield  {author} {\bibinfo {author} {\bibfnamefont {B.}~\bibnamefont
  {Butscher}}, \bibinfo {author} {\bibfnamefont {J.}~\bibnamefont {Nipper}},
  \bibinfo {author} {\bibfnamefont {J.~B.}\ \bibnamefont {Balewski}}, \bibinfo
  {author} {\bibfnamefont {L.}~\bibnamefont {Kukota}}, \bibinfo {author}
  {\bibfnamefont {V.}~\bibnamefont {Bendkowsky}}, \bibinfo {author}
  {\bibfnamefont {R.}~\bibnamefont {L\"ow}}, \ and\ \bibinfo {author}
  {\bibfnamefont {T.}~\bibnamefont {Pfau}},\ }\href {\doibase
  10.1038/nphys1828} {\bibfield  {journal} {\bibinfo  {journal} {Nat. Phys.}\
  }\textbf {\bibinfo {volume} {6}},\ \bibinfo {pages} {970} (\bibinfo {year}
  {2010})}\BibitemShut {NoStop}%
\bibitem [{\citenamefont {Li}\ \emph {et~al.}(2011)\citenamefont {Li},
  \citenamefont {Pohl}, \citenamefont {Rost}, \citenamefont {Rittenhouse},
  \citenamefont {Sadeghpour}, \citenamefont {Nipper}, \citenamefont {Butscher},
  \citenamefont {Balewski}, \citenamefont {Bendkowsky}, \citenamefont {L\"ow},\
  and\ \citenamefont {Pfau}}]{Li.2011}%
  \BibitemOpen
  \bibfield  {author} {\bibinfo {author} {\bibfnamefont {W.}~\bibnamefont
  {Li}}, \bibinfo {author} {\bibfnamefont {T.}~\bibnamefont {Pohl}}, \bibinfo
  {author} {\bibfnamefont {J.~M.}\ \bibnamefont {Rost}}, \bibinfo {author}
  {\bibfnamefont {S.~T.}\ \bibnamefont {Rittenhouse}}, \bibinfo {author}
  {\bibfnamefont {H.~R.}\ \bibnamefont {Sadeghpour}}, \bibinfo {author}
  {\bibfnamefont {J.}~\bibnamefont {Nipper}}, \bibinfo {author} {\bibfnamefont
  {B.}~\bibnamefont {Butscher}}, \bibinfo {author} {\bibfnamefont {J.~B.}\
  \bibnamefont {Balewski}}, \bibinfo {author} {\bibfnamefont {V.}~\bibnamefont
  {Bendkowsky}}, \bibinfo {author} {\bibfnamefont {R.}~\bibnamefont {L\"ow}}, \
  and\ \bibinfo {author} {\bibfnamefont {T.}~\bibnamefont {Pfau}},\ }\href
  {\doibase 10.1126/science.1211255} {\bibfield  {journal} {\bibinfo  {journal}
  {Science}\ }\textbf {\bibinfo {volume} {334}},\ \bibinfo {pages} {1110}
  (\bibinfo {year} {2011})}\BibitemShut {NoStop}%
\bibitem [{\citenamefont {Bendkowsky}\ \emph {et~al.}(2010)\citenamefont
  {Bendkowsky}, \citenamefont {Butscher}, \citenamefont {Nipper}, \citenamefont
  {Balewski}, \citenamefont {Shaffer}, \citenamefont {L\"ow}, \citenamefont
  {Pfau}, \citenamefont {Li}, \citenamefont {Stanojevic}, \citenamefont
  {Pohl},\ and\ \citenamefont {Rost}}]{Bendkowsky.2010}%
  \BibitemOpen
  \bibfield  {author} {\bibinfo {author} {\bibfnamefont {V.}~\bibnamefont
  {Bendkowsky}}, \bibinfo {author} {\bibfnamefont {B.}~\bibnamefont
  {Butscher}}, \bibinfo {author} {\bibfnamefont {J.}~\bibnamefont {Nipper}},
  \bibinfo {author} {\bibfnamefont {J.~B.}\ \bibnamefont {Balewski}}, \bibinfo
  {author} {\bibfnamefont {J.~P.}\ \bibnamefont {Shaffer}}, \bibinfo {author}
  {\bibfnamefont {R.}~\bibnamefont {L\"ow}}, \bibinfo {author} {\bibfnamefont
  {T.}~\bibnamefont {Pfau}}, \bibinfo {author} {\bibfnamefont {W.}~\bibnamefont
  {Li}}, \bibinfo {author} {\bibfnamefont {J.}~\bibnamefont {Stanojevic}},
  \bibinfo {author} {\bibfnamefont {T.}~\bibnamefont {Pohl}}, \ and\ \bibinfo
  {author} {\bibfnamefont {J.~M.}\ \bibnamefont {Rost}},\ }\href {\doibase
  10.1103/PhysRevLett.105.163201} {\bibfield  {journal} {\bibinfo  {journal}
  {Phys. Rev. Lett.}\ }\textbf {\bibinfo {volume} {105}},\ \bibinfo {pages}
  {163201} (\bibinfo {year} {2010})}\BibitemShut {NoStop}%
\bibitem [{\citenamefont {Gaj}\ \emph {et~al.}(2014)\citenamefont {Gaj},
  \citenamefont {Krupp}, \citenamefont {Balewski}, \citenamefont {L\"ow},
  \citenamefont {Hofferberth},\ and\ \citenamefont {Tilman}}]{Gaj.2014}%
  \BibitemOpen
  \bibfield  {author} {\bibinfo {author} {\bibfnamefont {A.}~\bibnamefont
  {Gaj}}, \bibinfo {author} {\bibfnamefont {A.~T.}\ \bibnamefont {Krupp}},
  \bibinfo {author} {\bibfnamefont {J.~B.}\ \bibnamefont {Balewski}}, \bibinfo
  {author} {\bibfnamefont {R.}~\bibnamefont {L\"ow}}, \bibinfo {author}
  {\bibfnamefont {S.}~\bibnamefont {Hofferberth}}, \ and\ \bibinfo {author}
  {\bibfnamefont {P.}~\bibnamefont {Tilman}},\ }\href@noop {} {\bibfield
  {journal} {\bibinfo  {journal} {Nat. Comm.}\ }\textbf {\bibinfo {volume} {5}}
  (\bibinfo {year} {2014})}\BibitemShut {NoStop}%
\bibitem [{\citenamefont {Fermi}(1934)}]{Fermi.1934}%
  \BibitemOpen
  \bibfield  {author} {\bibinfo {author} {\bibfnamefont {E.}~\bibnamefont
  {Fermi}},\ }\href {\doibase 10.1007/BF02959829} {\bibfield  {journal}
  {\bibinfo  {journal} {Il Nuovo Cimento}\ }\textbf {\bibinfo {volume} {11}},\
  \bibinfo {pages} {157} (\bibinfo {year} {1934})}\BibitemShut {NoStop}%
\bibitem [{\citenamefont {{Omont, A.}}(1977)}]{Omont.1977}%
  \BibitemOpen
  \bibfield  {author} {\bibinfo {author} {\bibnamefont {{Omont, A.}}},\ }\href
  {\doibase 10.1051/jphys:0197700380110134300} {\bibfield  {journal} {\bibinfo
  {journal} {J. Phys. France}\ }\textbf {\bibinfo {volume} {38}},\ \bibinfo
  {pages} {1343} (\bibinfo {year} {1977})}\BibitemShut {NoStop}%
\bibitem [{\citenamefont {Greene}\ \emph {et~al.}(2000)\citenamefont {Greene},
  \citenamefont {Dickinson},\ and\ \citenamefont {Sadeghpour}}]{Greene.2000}%
  \BibitemOpen
  \bibfield  {author} {\bibinfo {author} {\bibfnamefont {C.~H.}\ \bibnamefont
  {Greene}}, \bibinfo {author} {\bibfnamefont {A.~S.}\ \bibnamefont
  {Dickinson}}, \ and\ \bibinfo {author} {\bibfnamefont {H.~R.}\ \bibnamefont
  {Sadeghpour}},\ }\href {\doibase 10.1103/PhysRevLett.85.2458} {\bibfield
  {journal} {\bibinfo  {journal} {Phys. Rev. Lett.}\ }\textbf {\bibinfo
  {volume} {85}},\ \bibinfo {pages} {2458} (\bibinfo {year}
  {2000})}\BibitemShut {NoStop}%
\bibitem [{\citenamefont {Fabrikant}(1986)}]{Fabrikant.1986}%
  \BibitemOpen
  \bibfield  {author} {\bibinfo {author} {\bibfnamefont {I.~I.}\ \bibnamefont
  {Fabrikant}},\ }\href {http://stacks.iop.org/0022-3700/19/i=10/a=021}
  {\bibfield  {journal} {\bibinfo  {journal} {J. Phys. B.}\ }\textbf {\bibinfo
  {volume} {19}},\ \bibinfo {pages} {1527} (\bibinfo {year}
  {1986})}\BibitemShut {NoStop}%
\bibitem [{\citenamefont {Bahrim}\ and\ \citenamefont
  {Thumm}(2000)}]{Bahrim.2000}%
  \BibitemOpen
  \bibfield  {author} {\bibinfo {author} {\bibfnamefont {C.}~\bibnamefont
  {Bahrim}}\ and\ \bibinfo {author} {\bibfnamefont {U.}~\bibnamefont {Thumm}},\
  }\href {\doibase 10.1103/PhysRevA.61.022722} {\bibfield  {journal} {\bibinfo
  {journal} {Phys. Rev. A}\ }\textbf {\bibinfo {volume} {61}},\ \bibinfo
  {pages} {022722} (\bibinfo {year} {2000})}\BibitemShut {NoStop}%
\bibitem [{\citenamefont {Bahrim}\ \emph {et~al.}(2001)\citenamefont {Bahrim},
  \citenamefont {Thumm},\ and\ \citenamefont {Fabrikant}}]{Bahrim.2001}%
  \BibitemOpen
  \bibfield  {author} {\bibinfo {author} {\bibfnamefont {C.}~\bibnamefont
  {Bahrim}}, \bibinfo {author} {\bibfnamefont {U.}~\bibnamefont {Thumm}}, \
  and\ \bibinfo {author} {\bibfnamefont {I.~I.}\ \bibnamefont {Fabrikant}},\
  }\href {http://stacks.iop.org/0953-4075/34/i=6/a=107} {\bibfield  {journal}
  {\bibinfo  {journal} {Journal of Physics B: Atomic, Molecular and Optical
  Physics}\ }\textbf {\bibinfo {volume} {34}},\ \bibinfo {pages} {L195}
  (\bibinfo {year} {2001})}\BibitemShut {NoStop}%
\bibitem [{\citenamefont {Hamilton}\ \emph {et~al.}(2002)\citenamefont
  {Hamilton}, \citenamefont {Greene},\ and\ \citenamefont
  {Sadeghpour}}]{Hamilton.2002}%
  \BibitemOpen
  \bibfield  {author} {\bibinfo {author} {\bibfnamefont {E.~L.}\ \bibnamefont
  {Hamilton}}, \bibinfo {author} {\bibfnamefont {C.~H.}\ \bibnamefont
  {Greene}}, \ and\ \bibinfo {author} {\bibfnamefont {H.~R.}\ \bibnamefont
  {Sadeghpour}},\ }\href {http://stacks.iop.org/0953-4075/35/i=10/a=102}
  {\bibfield  {journal} {\bibinfo  {journal} {Journal of Physics B: Atomic,
  Molecular and Optical Physics}\ }\textbf {\bibinfo {volume} {35}},\ \bibinfo
  {pages} {L199} (\bibinfo {year} {2002})}\BibitemShut {NoStop}%
\bibitem [{\citenamefont {Khuskivadze}\ \emph {et~al.}(2002)\citenamefont
  {Khuskivadze}, \citenamefont {Chibisov},\ and\ \citenamefont
  {Fabrikant}}]{Khuskivadze.2002}%
  \BibitemOpen
  \bibfield  {author} {\bibinfo {author} {\bibfnamefont {A.~A.}\ \bibnamefont
  {Khuskivadze}}, \bibinfo {author} {\bibfnamefont {M.~I.}\ \bibnamefont
  {Chibisov}}, \ and\ \bibinfo {author} {\bibfnamefont {I.~I.}\ \bibnamefont
  {Fabrikant}},\ }\href {\doibase 10.1103/PhysRevA.66.042709} {\bibfield
  {journal} {\bibinfo  {journal} {Phys. Rev. A}\ }\textbf {\bibinfo {volume}
  {66}},\ \bibinfo {pages} {042709} (\bibinfo {year} {2002})}\BibitemShut
  {NoStop}%
\bibitem [{\citenamefont {Chibisov}\ \emph {et~al.}(2002)\citenamefont
  {Chibisov}, \citenamefont {Khuskivadze},\ and\ \citenamefont
  {Fabrikant}}]{Chibisov.2002}%
  \BibitemOpen
  \bibfield  {author} {\bibinfo {author} {\bibfnamefont {M.~I.}\ \bibnamefont
  {Chibisov}}, \bibinfo {author} {\bibfnamefont {A.~A.}\ \bibnamefont
  {Khuskivadze}}, \ and\ \bibinfo {author} {\bibfnamefont {I.~I.}\ \bibnamefont
  {Fabrikant}},\ }\href {http://stacks.iop.org/0953-4075/35/i=10/a=101}
  {\bibfield  {journal} {\bibinfo  {journal} {Journal of Physics B: Atomic,
  Molecular and Optical Physics}\ }\textbf {\bibinfo {volume} {35}},\ \bibinfo
  {pages} {L193} (\bibinfo {year} {2002})}\BibitemShut {NoStop}%
\bibitem [{\citenamefont {Meschede}(1987)}]{Meschede1987}%
  \BibitemOpen
  \bibfield  {author} {\bibinfo {author} {\bibfnamefont {D.}~\bibnamefont
  {Meschede}},\ }\href@noop {} {\bibfield  {journal} {\bibinfo  {journal} {J.
  Opt. Soc. Am. B}\ }\textbf {\bibinfo {volume} {4}},\ \bibinfo {pages} {413 }
  (\bibinfo {year} {1987})}\BibitemShut {NoStop}%
\bibitem [{\citenamefont {Tauschinsky}\ \emph {et~al.}(2013)\citenamefont
  {Tauschinsky}, \citenamefont {Newell}, \citenamefont {van Linden van~den
  Heuvell},\ and\ \citenamefont {Spreeuw}}]{Tauschinsky2013}%
  \BibitemOpen
  \bibfield  {author} {\bibinfo {author} {\bibfnamefont {A.}~\bibnamefont
  {Tauschinsky}}, \bibinfo {author} {\bibfnamefont {R.}~\bibnamefont {Newell}},
  \bibinfo {author} {\bibfnamefont {H.~B.}~\bibnamefont {van~Linden~van~den~Heuvell}}, \ and\ \bibinfo {author} {\bibfnamefont {R.~J.~C.}~\bibnamefont
  {Spreeuw}},\ }\href@noop {} {\bibfield  {journal} {\bibinfo  {journal} {Phys.
  Rev. A}\ }\textbf {\bibinfo {volume} {87}},\ \bibinfo {pages} {042522 (5 pp.)
  } (\bibinfo {year} {2013})}\BibitemShut {NoStop}%
\bibitem [{\citenamefont {Sassmannshausen}\ \emph {et~al.}(2013)\citenamefont
  {Sassmannshausen}, \citenamefont {Merkt},\ and\ \citenamefont
  {Deiglmayr}}]{Sassmannshausen2013}%
  \BibitemOpen
  \bibfield  {author} {\bibinfo {author} {\bibfnamefont {H.}~\bibnamefont
  {Sassmannshausen}}, \bibinfo {author} {\bibfnamefont {F.}~\bibnamefont
  {Merkt}}, \ and\ \bibinfo {author} {\bibfnamefont {J.}~\bibnamefont
  {Deiglmayr}},\ }\href@noop {} {\bibfield  {journal} {\bibinfo  {journal}
  {Phys. Rev. A}\ }\textbf {\bibinfo {volume} {87}},\ \bibinfo {pages} {032519
  (10 pp.) } (\bibinfo {year} {2013})}\BibitemShut {NoStop}%
\bibitem [{\citenamefont {T.F.Gallagher}(1994)}]{Gallagher}%
  \BibitemOpen
  \bibfield  {author} {\bibinfo {author} {\bibnamefont {T.F.Gallagher}},\
  }\href@noop {} {\emph {\bibinfo {title} {Rydberg Atoms}}}\ (\bibinfo
  {publisher} {Cambridge University Press},\ \bibinfo {address} {New York, NY,
  USA},\ \bibinfo {year} {1994})\ p.\ \bibinfo {pages} {349}\BibitemShut
  {NoStop}%
\bibitem [{\citenamefont {Litzén}(1970)}]{Litzen.1970}%
  \BibitemOpen
  \bibfield  {author} {\bibinfo {author} {\bibfnamefont {U.}~\bibnamefont
  {Litzén}},\ }\href {http://iopscience.iop.org/1402-4896/1/5-6/012}
  {\bibfield  {journal} {\bibinfo  {journal} {Physica Scripta}\ }\textbf
  {\bibinfo {volume} {1}},\ \bibinfo {pages} {253} (\bibinfo {year}
  {1970})}\BibitemShut {NoStop}%
\bibitem [{\citenamefont {Sadeghpour}\ and\ \citenamefont
  {Rittenhouse}(2013)}]{Sadeghpour.2013}%
  \BibitemOpen
  \bibfield  {author} {\bibinfo {author} {\bibfnamefont {H.~R.}\ \bibnamefont
  {Sadeghpour}}\ and\ \bibinfo {author} {\bibfnamefont {S.~T.}\ \bibnamefont
  {Rittenhouse}},\ }\href {\doibase 10.1080/00268976.2013.811555} {\bibfield
  {journal} {\bibinfo  {journal} {Molecular Physics}\ }\textbf {\bibinfo
  {volume} {111}},\ \bibinfo {pages} {1902} (\bibinfo {year}
  {2013})}\BibitemShut {NoStop}%
\bibitem [{\citenamefont {Sakurai}(1994)}]{Sakurai.Modern}%
  \BibitemOpen
  \bibfield  {author} {\bibinfo {author} {\bibfnamefont {J.~J.}\ \bibnamefont
  {Sakurai}},\ }\href@noop {} {\emph {\bibinfo {title} {Modern Quantum
  Mechanics Revised Edition}}}\ (\bibinfo {year} {1994})\BibitemShut {NoStop}%
\bibitem [{\citenamefont {Wigner}(1955)}]{Wigner.1955}%
  \BibitemOpen
  \bibfield  {author} {\bibinfo {author} {\bibfnamefont {E.~P.}\ \bibnamefont
  {Wigner}},\ }\href {\doibase 10.1103/PhysRev.98.145} {\bibfield  {journal}
  {\bibinfo  {journal} {Phys. Rev.}\ }\textbf {\bibinfo {volume} {98}},\
  \bibinfo {pages} {145} (\bibinfo {year} {1955})}\BibitemShut {NoStop}%
\bibitem [{\citenamefont {Butscher}\ \emph {et~al.}(2011)\citenamefont
  {Butscher}, \citenamefont {Bendkowsky}, \citenamefont {Nipper}, \citenamefont
  {Balewski}, \citenamefont {Kukota}, \citenamefont {Löw}, \citenamefont
  {Pfau}, \citenamefont {Li}, \citenamefont {Pohl},\ and\ \citenamefont
  {Rost}}]{Butscher.2011}%
  \BibitemOpen
  \bibfield  {author} {\bibinfo {author} {\bibfnamefont {B.}~\bibnamefont
  {Butscher}}, \bibinfo {author} {\bibfnamefont {V.}~\bibnamefont
  {Bendkowsky}}, \bibinfo {author} {\bibfnamefont {J.}~\bibnamefont {Nipper}},
  \bibinfo {author} {\bibfnamefont {J.~B.}\ \bibnamefont {Balewski}}, \bibinfo
  {author} {\bibfnamefont {L.}~\bibnamefont {Kukota}}, \bibinfo {author}
  {\bibfnamefont {R.}~\bibnamefont {Löw}}, \bibinfo {author} {\bibfnamefont
  {T.}~\bibnamefont {Pfau}}, \bibinfo {author} {\bibfnamefont {W.}~\bibnamefont
  {Li}}, \bibinfo {author} {\bibfnamefont {T.}~\bibnamefont {Pohl}}, \ and\
  \bibinfo {author} {\bibfnamefont {J.~M.}\ \bibnamefont {Rost}},\ }\href
  {http://stacks.iop.org/0953-4075/44/i=18/a=184004} {\bibfield  {journal}
  {\bibinfo  {journal} {Journal of Physics B: Atomic, Molecular and Optical
  Physics}\ }\textbf {\bibinfo {volume} {44}},\ \bibinfo {pages} {184004}
  (\bibinfo {year} {2011})}\BibitemShut {NoStop}%
\bibitem [{\citenamefont {Junginger}\ \emph {et~al.}(2013)\citenamefont
  {Junginger}, \citenamefont {Main},\ and\ \citenamefont
  {Wunner}}]{Junginger.2013}%
  \BibitemOpen
  \bibfield  {author} {\bibinfo {author} {\bibfnamefont {A.}~\bibnamefont
  {Junginger}}, \bibinfo {author} {\bibfnamefont {J.}~\bibnamefont {Main}}, \
  and\ \bibinfo {author} {\bibfnamefont {G.}~\bibnamefont {Wunner}},\ }\href
  {http://stacks.iop.org/0953-4075/46/i=8/a=085201} {\bibfield  {journal}
  {\bibinfo  {journal} {Journal of Physics B: Atomic, Molecular and Optical
  Physics}\ }\textbf {\bibinfo {volume} {46}},\ \bibinfo {pages} {085201}
  (\bibinfo {year} {2013})}\BibitemShut {NoStop}%
\bibitem [{\citenamefont {Brown}\ and\ \citenamefont
  {Carrington}(2003)}]{Brown}%
  \BibitemOpen
  \bibfield  {author} {\bibinfo {author} {\bibfnamefont {J.~M.}\ \bibnamefont
  {Brown}}\ and\ \bibinfo {author} {\bibfnamefont {A.}~\bibnamefont
  {Carrington}},\ }\href@noop {} {\emph {\bibinfo {title} {Rotational
  spectroscopy of diatomic molecules}}}\ (\bibinfo {year} {2003})\BibitemShut
  {NoStop}%
\end{thebibliography}

\end{document}